\documentstyle[prd,aps,floats,epsfig]{revtex}
\tighten

\begin{document}
\draft
\title{Fermionic current-carrying cosmic strings: \\
zero-temperature limit and equation of state}
\author{P.~Peter\thanks{\tt peter@iap.fr},
and C.~Ringeval\thanks{\tt ringeval@iap.fr} }
\address{Institut d'Astrophysique de Paris, 98bis boulevard
Arago, 75014 Paris, France}

\maketitle

\begin{abstract}
The equation of state for a superconducting cosmic string whose
current is due to fermionic zero modes is derived analytically in the
case where the back-reaction of the fermions to the background is
neglected. It is first shown that the zero mode fermions follow a zero
temperature distribution because of their interactions (or lack
thereof) with the string-forming Higgs and gauge fields. It is then
found that the energy per unit length $U$ and the tension $T$ are
related to the background string mass $m$ through the simple relation
$U+T= 2m^2$.  Cosmological consequences are briefly discussed.
\end{abstract}



\section{Introduction}

Topological defects~\cite{NATO} have been considered in various
physical situations, e.g., in the context of condensed matter and
cosmology~\cite{kibble,book}. In many cases of interest in
cosmology~\cite{simulations,BPRS}, they can be approximated as
structureless, the relevant dynamics being often assumed not to
depend on any specific choice of their internal content. For
cosmic defects, this internal content would correspond to the
particles that couple to the string-forming Higgs
field~\cite{witten}. However, in the latter example of cosmic
strings, it was shown that such a structure might lead to drastic
modifications not only of these object
dynamics~\cite{formal,neutral}, which could be seen as a mere
academic situation given our present ignorance on their very
existence, but also, because of the appearance of new accessible
equilibrium states, of the cosmological setting, leading in some
instances to actual catastrophes~\cite{vortons,rdp}. To make a
long story short, let us just say that currents imply a breakdown
of the Lorentz-boost invariance along the string worldsheet,
thereby allowing loop configurations to rotate, the centrifugal
force hereby induced having the ability to sustain the loop
tendency to shrink because of the tension. The resulting states,
called vortons, might be stable even over cosmological timescales,
scaling as matter and thus rapidly coming to dominate the
Universe evolution, in contradiction with the observations. This
leads to constraints on the particle physics theories that
predict them at energy scales that are believed to be unreachable
experimentally (in accelerators say) in the foreseeable future.

Unfortunately, it appears that the string structure, contrary to
their counterparts as fundamental objects~\cite{superstring}, is
not determined by any consistency relation, and is therefore
somehow arbitrary, at least at the effective description
level~\cite{witten}. This means in practice that in order to be
able to tell anything relevant to (cosmic) string cosmology, one
needs to set up a complete underlying microscopic model, arising
say, from ones favorite Grand Unified Theory (GUT)~\cite{davis}
or some low-energy approximation of some superstring-inspired
model~\cite{dbp}.

Some generic constructions can however be arranged, as was shown
to be the case whence a bosonic condensate gets frozen in the
string core~\cite{formal,neutral}. In such a situation, the boson
field phase $\varphi$, thanks to a random Kibble-like mechanism,
may wind along the string itself, thereby producing a current that
turns out to be essentially a function of a single state
parameter $w$, thus expressible as a phase gradient as
\begin{equation} w \equiv \kappa_0 \gamma^{ab} \partial_a \varphi
\partial _b \varphi,\label{w}\end{equation}
with indices $a,b,...$ varying within the string worldsheet
coordinates defined by the relations
\begin{equation} x^\mu = X^\mu_{_{S}} (\xi^a), \ \ \ \ \ \
\xi^a \in \{\tau,\sigma\},\end{equation}
and $\gamma^{ab}$ the inverse of the induced metric defined with the
background metric $g_{\mu\nu}$ as
\begin{equation}\gamma_{ab} = g_{\mu\nu} {\partial
X^\mu_{_{S}} \over \partial \xi^a} {\partial X^\nu_{_{S}} \over
\partial \xi^b}.\label{gammaab}\end{equation} A straightforward
generalization of the Nambu-Goto action is then provided by the
$w-$weighted measure as
\begin{equation} {\cal S} = -m^2 \int \hbox{d}^2\xi \sqrt{-\gamma}
{\cal L} \{w\},\end{equation} with $m$ the typical mass scale of
symmetry breaking leading to string formation and $\gamma$ the
determinant of the induced metric~(\ref{gammaab}).  Reasonable
microscopic models~\cite{carpet1} then yield approximate forms for
the Lagrangian function ${\cal L}\{w\}$, out of which the
dynamical properties of the corresponding strings can be
derived~\cite{larsen,cpg,gpb}.

Such a description remains however essentially classical even
though an alternative formalism, also proposed by
Carter~\cite{sigma}, in terms of a dilatonic model, appears more
suitable for quantization. This last formalism however, being
fully two-dimensional, cannot be used to derive interesting
quantities such as the relevant cross-sections for trapped
excitations to leave the string worldsheet. This is unfortunate
since this is precisely the information one would need for
cosmological applications~\cite{vortons,davis}.

It would therefore seem that by considering fermionic current
carriers instead of bosonic ones, one would, because of the
intrinsically quantum nature of fermions, obtain a more
appropriate description~\cite{davis,chris}. Besides, fermions are
trapped in topological defects because of Yukawa couplings with
the string forming Higgs field in the form of zero
modes~\cite{Jackiw}, so that their dynamics is described by
simple (although coupled) Dirac equations, which are linear. In
the bosonic case, the non-linear (quartic) term is essential in
order to ensure the dynamical stability of the condensate so that
a solitonic treatment~\cite{soliton,hp} seems the only way to
deal with the underlying quantum physics. This fact dramatically
complicates matters and as a result, a complete description yet
fails to exist.

However, the fermionic case is not that simple either as here, one
faces another technical difficulty for the classical description:
it can be shown that there doesn't exist a simple state
parameter~\cite{chris}. An arbitrary spacelike or timelike
current can only be built out of at least two opposite chirality
spinor fields and will be given by the knowledge of four
occupation numbers per unit length. As this is true in particular
at least in the zero-temperature limit, we shall be concerned
here first with this limit whose validity was assumed to depend
on the particular model under consideration~\cite{davis}. In the
following section, we show that setting the temperature to zero
is always a good approximation because of the couplings between
the fermionic fields and the string-forming Higgs and gauge
fields\footnote{Otherwise, the temperature itself could be taken
as a state parameter, so that the usual formalism would be
applicable~\cite{cartertemp}}. These results would seem to imply
that the previously derived macroscopic formalism is irrelevant
to the fermionic situation (see Ref.~\cite{many} for a
many-parameter formalism). In practice however, as a simple
relationship between the energy per unit length and the tension
can be found for fermionic currents, the single state parameter
formalism can be used that permits to draw some cosmological
consequences.

\section{The zero-temperature limit}

In order to have an arbitrary current built upon fermionic fields,
one needs at least two Dirac fermions~\cite{witten} $\Psi$ and
$\chi$ say, coupled to the string-forming Higgs field $\Phi$
through Yukawa terms as well as to the associated gauge field
$B_\mu$, the later acquiring a mass from the vacuum expectation
value (VEV) of the Higgs field. Fermions may condense in the
string core in the form of zero modes, and by filling up the
accessible states, one forms a current which can be timelike,
spacelike, or lightlike. If one wants to give a classical
description of such a current-carrying string, one must be in a
configuration for which quantum effects are negligible. Such
quantum effects, as for instance tunneling outside the vortex,
will indeed be negligible provided most of the fermions are on
energy levels whose excitation energy is much below their vacuum
mass, the latter thus playing the role of a Fermi energy.

As a result, if a temperature may be defined for the fermion
ensemble along the vortex, quantum effects will be negligible if
the temperature is small compared to the vacuum mass of the
fermions, which essentially imply a zero temperature state. Note
that the situation we are having in mind is reached only whenever
the background temperature is low compared to that at which the
string formed for otherwise interactions with the surrounding
plasma could populate the high energy levels. In practice, this is
what will happen at the time of string formation, and if the
fermions did not interact at all, or only through time reversible
interactions, one would be left with a frozen distribution
corresponding to a high temperature state~\cite{bcpc}.

That this is not the case can be seen through an exhaustive list
of all the possible fermion interactions. For that purpose, it
must be emphasized that fermionic condensates arise in the string
core in the form of zero modes, i.e., chiral states\footnote{We
do not consider here the possible massive bound states as those
are expected~\cite{davis} to interact with each other because of
diagram $(c)$ of the figure and therefore move rapidly away from
the string core.} . One has essentially two coupling
possibilities, namely a coupling of the fermion with the Higgs
field or with the gauge field, illustrated on the figure. The
first case (diagram $a$) is seen to vanish identically in the
case of a chiral mode~\cite{chris} so we shall not consider it.
The second case is more interesting and comes from the second
diagram ($b$) of the figure. In this case, any trapped fermionic
zero mode is seen to be able to radiate a gauge vector boson.
These terms do not vanish identically, and in fact can be seen to
be the source for some backreacted components of the gauge field:
as all the vectors emitted this way will eventually condense into
a classical field, this diagram contributes to a small (and
indeed usually negligible~\cite{chris}) contribution in the
energy per unit length and tension.

\begin{figure}[h]
\begin{center}
\epsfig{file=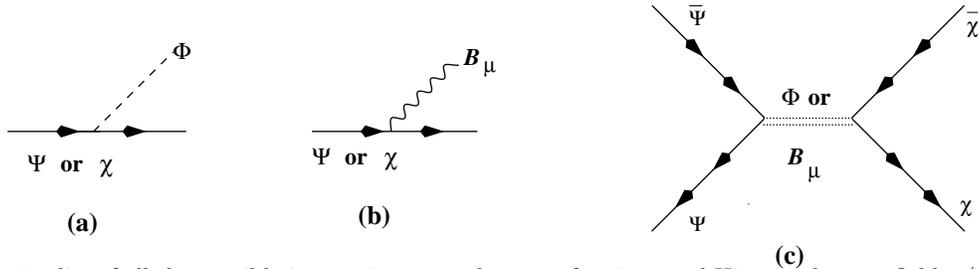,width=13cm}
\caption{Exhaustive
list of all the possible interaction terms between fermions and
Higgs and gauge fields. $(a)$ Higgs radiation by a fermion field
denoted by $\psi$ or $\chi$ (two fermion at least are necessary
in order to generate an arbitrary kind of current), $(b)$ gauge
boson radiation, $(c)$ coupling between the various fermion
fields.}
\end{center}
\end{figure}

Finally the third term, $(c)$, of the figure, represents a
would-be interaction term between the fermions that could be
responsible for an equilibrium configuration. When one considers
only zero modes in a usual model, such a term actually vanishes
because both fermions must have opposite chiralities in order for
the theory to be well defined. As a result, interactions between
fermions turn out to be negligible.

We are now in a position to understand the microphysics of what
might happen inside a fermionic current-carrying cosmic string.
First, when the fermions become trapped in the string core, they
do so on arbitrary high energy levels inside the string. Then,
they have the possibility to radiate most of their energy away in
the form of the vector field, thereby creating the backreacted
component. Note that the vector field itself can also interact
with the Higgs field thereby producing an effective lack of
symmetry between diagram $(b)$ and its time reversed counterpart.
This implies that the overall effect is indeed a radiative decay
and not an equilibrium. Thus, all the populated states end up
being the lowest reachable states. In practice, that means that
the effective temperature of the fermion gas is vanishing.
Moreover, such a configuration in turn is stable as no
interaction between the various fermion field can be present.

\section{Fermionic string equation of state}

As was discussed above, fermions that are coupled to the Higgs
field may be bound to cosmic strings in the form of zero
modes~\cite{witten,chris,Jackiw} and therefore the current they
generate arises from lightlike components. We recall briefly the
formalism necessary to handle this case and then move on to
derive the resulting stress-energy tensor as well as a simple
relationship relating its eigenvalues. This leads to a plausible
classical description in terms of a state parameter whose
validity is discussed.

\subsection{Stress-energy tensor and equation of state}

Chiral currents have special properties and need be studied on
their own. As was shown earlier, a phase gradient formalism hold
for them, similar to the $\sigma-$model~\cite{sigma} with
vanishing potential, namely~\cite{carpet2}
\begin{equation} {\cal S}_\sigma = - \int \hbox{d}^2\xi \sqrt{-\gamma}
\left( m^2 +{1\over 2} \psi^2 \gamma^{ab} \partial_a\varphi \partial_b
\varphi \right),\label{chiral}\end{equation}
where the normalization constant $\kappa_0$ of Eq.~(\ref{w}) in the
$w-$formalism has now been promoted to a dynamical field whose
variations lead to an ever-lightlike current.

Considering now a system of fermionic zero modes, and neglecting
back-reaction, one may arrive at the conclusion that the relevant
action describing a general fermionic current-carrier cosmic string
will be given by
\begin{equation} {\cal S}_{_F} = \int \hbox{d}^2\xi \sqrt{-\gamma}
{\cal L} \end{equation}
where the Lagrangian function ${\cal L}$ is
\begin{equation}{\cal L} =-m^2 -{1\over 2} \sum_i^N \psi^2_{(i)} \gamma^{ab}
\partial_a\varphi_{(i)} \partial_b
\varphi_{(i)},\label{manych}\end{equation} $N$ being the number of
fermionic degrees of freedom (at least four~\cite{chris} if the
model is to describe arbitrary spacelike as well as timelike and
chiral currents). Having obtained the action, it is now a simple
matter to derive the corresponding dynamics by varying it with
respect to the various fields involved. However, as we show
below, it turns out not to be strictly necessary as many
consequences, in particular in cosmology, stem directly from the
energy momentum tensor eigenvalues $U$ and $T$, i.e.,
respectively the energy per unit length and tension, for which a
very simple relationship is now derived.

First of all, as all the fields $\psi_{(i)}$ are independent, it is
evident that variations of (\ref{manych}) lead to the chirality
condition on the various currents induced by the phase gradients,
namely
\begin{equation} {\delta {\cal S}_{F}\over \delta \psi_{(i)}} = 0 \ \
\Longrightarrow \ \ \ \gamma^{ab} \partial_a\varphi_{(i)}
\partial_b \varphi_{(i)} = 0 \ \ \ \forall i\in
[1,N],\label{dfdf}\end{equation} while the currents are obtained
through variations with respect to the phases themselves
\begin{equation} {\delta {\cal S}_{F}\over \delta \varphi_{(i)}} = 0 \ \
\Longrightarrow \ \ \ \nabla_a \left(\psi^2_{(i)} \gamma^{ab}
\partial_b \varphi_{(i)}\right) \equiv \nabla_a j^a_{(i)} =
0, \ \ \ \forall i\in [1,N].\label{curr}\end{equation}
Eq.~(\ref{dfdf}) can be seen to imply, in our two-dimensional
case, that each function $\varphi_{(i)}(\xi^a)$ separately is
harmonic, i.e.
\begin{equation} \gamma^{ab} \nabla_a \nabla_b \varphi_{(i)} = 0\ \ \
\forall i\in [1,N],\label{harm}\end{equation} so that the current
conservation equations (\ref{curr}) can be cast in the form
\begin{equation} \gamma^{ab} \nabla_a \psi_{(i)} \nabla_b
\varphi_{(i)} = 0\ \ \ \forall i\in [1,N],\end{equation} which
means that every $\psi_{(i)}$ is a function of $\varphi_{(i)}$
only, for any fixed value of $i$. As a result, the formalism
really describes only $N$ degrees of freedom, and not $2N$, and
one may interpret the phase gradients as occupation numbers per
unit length in a given underlying fermionic current-carrying
model.

The stress energy tensor is now obtained by the standard procedure of
variation with respect to the metric, i.e.,
\begin{equation} \overline T^{\mu\nu} = 2{\delta {\cal L}\over
\delta g_{\mu\nu}} +{\cal L} \eta^{\mu\nu},\end{equation} where
the first fundamental tensor of the worldsheet~\cite{formal}
\begin{equation} \eta^{\mu\nu} = \gamma^{ab} {\partial
X^\mu_{_{S}} \over \partial \xi^a} {\partial X^\nu_{_{S}} \over
\partial \xi^b}\label{eta}\end{equation} is definable in terms of the
eigenvectors $u^\mu$ and $v^\mu$, respectively timelike and
spacelike ($u_\mu u^\mu = - v_\mu v^\nu = -1$), of the stress
energy tensor for a non-chiral current:
\begin{equation} \eta^{\mu\nu} = v^\mu v^\nu - u^\mu u^\nu,
\end{equation}
and
\begin{equation} \overline T^{\mu\nu} = U u^\mu u^\nu - T v^\mu v^\nu
\label{UT}.\end{equation}

The stress-energy tensor now reads
\begin{equation} \overline T^{\mu\nu} = - m^2\eta^{\mu\nu}
+{1\over 2} \sum_i^N \psi_{(i)}^2 \left( X^\mu_{_{S},c}
X^\nu_{_{S},d} \varphi_{(i)} ^{,c}\varphi_{(i)} ^{,d} -
\eta^{\mu\nu} \varphi_{(i),a} \varphi_{(i)}
^{,a}\right)\label{stress}\end{equation} with a comma denoting
partial differentiation with respect to a worldsheet coordinate.
The last term identically vanishes because of the on-shell
relation (\ref{dfdf}), and we can now compute the eigenvalues by
projecting on the eigenvectors as
\begin{equation} U = \overline T^{\mu\nu} u_\mu u_\nu = m^2 + {1\over 2}
\left( \sum_i^N \psi_{(i)}^2 \varphi_{(i)}^{,a} \varphi_{(i)}^{,b}
\right) X^\mu_{_{S},a} X^\nu_{_{S},b} u_\mu u_\nu,
\end{equation}
and
\begin{equation} T = - \overline T^{\mu\nu} v_\mu v_\nu
= m^2 - {1\over 2} \left( \sum_i^N \psi_{(i)}^2
\varphi_{(i)}^{,a} \varphi_{(i)}^{,b} \right) X^\mu_{_{S},a}
X^\nu_{_{S},b} v_\mu v_\nu.
\end{equation}
By adding these two equations up, one gets that the last term is
proportional to the first fundamental tensor $\eta^{\mu\nu}$
projected onto the worldsheet coordinates, i.e. a term
proportional to the induced metric $\gamma_{ab}$. As each part of
the current is made of chiral fields, this last term eventually
cancels out and one is left with [a relation obtainable directly
by taking the trace of the stress tensor~(\ref{stress})]
\begin{equation} U+T = 2m^2,\label{UTm2}\end{equation}
which will be our final equation of state for a fermionic current
carrying cosmic string in the zero temperature limit. Note that
this relation holds in the specific case of the model discussed
in Ref.~\cite{chris} whenever one neglects the fermion
backreaction on the string fields.

\subsection{A macroscopic model}

Let us now discuss the various implications of this result. The
most important point related with cosmological models involving
current-carrying strings concerns vorton stability. As such a
model is exclusively classical in nature, we shall not examine the
quantum stability here, especially since this was already
discussed in Ref.~\cite{davis}. Before however turning to this
physical point, we should like to stress a simple technical detail
concerning the equation of state itself.

As we have said, a fermionic current-carrying cosmic string does
not in general admit a classical description in terms of a single
state parameter. However, in the case where a functional
relationship exists between the energy per unit length and the
tension, as is indeed what happens in the situation under
consideration here, a state parameter can easily be derived as
follows.

Let us consider again the $w-$formalism. Performing the Legendre
transform
\begin{equation} \Lambda = {\cal L} - 2w{\hbox{d}{\cal
L}\over\hbox{d} w},\label{Legendre} \end{equation} it can be
shown that~\cite{formal}, depending on the timelike or spacelike
character of the current, the energy per unit length and tension
can be identified, up to a sign, with ${\cal L}$ and $\Lambda$.
As a result, the knowledge of ${\cal L}$ as a function of
$\Lambda$ or, in other words that of $U(T)$, permits to integrate
Eq.~(\ref{Legendre}) to yield the state parameter through
\begin{equation}
\ln \left({w\over w_0}\right) = \int {\hbox{d}{\cal L}\over
2({\cal L} - \Lambda)},\label{state}
\end{equation} whose inversion, in turn, gives the functional
form of the Lagrangian ${\cal L}\{w\}$, up to a normalization
factor. Applied to our case, Eq.~(\ref{state}) implies immediately
\begin{equation} {\cal L}\{w\} = -m^2 -{w\over 2},
\label{linear}\end{equation} so we see that an arbitrary current
formed with many lightlike currents can be described by means of
a single state parameter with almost the simplest possible model;
in Eq.~(\ref{manych}), it suffices to replace the sum over the
many chiral models by the standard form of $w$, i.e.
Eq.~(\ref{w}), which can be viewed as the auxiliary field $\psi$
acquiring a fixed value ''on shell''. This is just the action of
Eq.~(\ref{chiral}) with $\psi^2 = \kappa_0$.

The model described by Eq.~(\ref{linear}) was however ruled out
as a valid description of a realistic Witten-like
current-carrying string in Ref.~\cite{carpet1}, so one may wonder
how it can be re-introduced here. There are two answers to that
question. First, it can be argued that most of the statements in
this reference applied to bosonic current-carriers, and have
therefore no reason to be true in the fermionic case, except that
bosons and fermions are known to be equivalent in two
dimensions~\cite{witten}. As a result, a classical description of
a vortex must somehow take into account the finite thickness
effects before averaging over the transverse degrees of freedom,
so that the string keeps a track of its $3+1$-dimensional nature.

The second, perhaps more important reason, why the model given by
Eq.~(\ref{linear}) was not considered seriously as a candidate to
describe a current-carrying string is the saturation effect. There
must indeed be a maximum current flowing along a string as individual
particles making the current are limited in energy because they are
bound states. In the case of bosons, thanks to Bose condensate, all
the particles are essentially in the same state and the saturation
effect stems from the non-linear (interaction) term between them. As
it turns out, even the interaction terms can be adequately treated
through a mean field approximation, so that a classical field
description is valid in this case. For fermions however, this effect
finds its origin in a completely different mechanism, related to the
exclusion principle: it is necessary, in order to increase the value
of the current, to add more particles on higher and higher energy
levels, up to the point where it becomes energetically favorable for
them to leave the worldsheet as massive modes. This is therefore a
purely quantum effect which cannot, of course, be properly taken into
account in the classical description developed here whose range of
validity is thus limited to small currents. Moreover, contrary to the
bosonic situation in which the boson mass enters explicitly as a
relevant dynamical parameter, fermionic zero modes exist independently
of the vacuum fermion mass, so there is no mass scale that could
determine the saturation regime in such a classical description.

The conclusion of the previous discussion is that the model described
by Eq.~(\ref{linear}) is indeed an accurate representation for
fermionic current-carrying cosmic strings provided the current is far
from the saturation regime. It should be emphasized that it will be
the case for most of the evolution of a network of such strings, so
that one is entitled, for cosmological application purposes
(e.g. numerical simulation), to derive the string dynamics with the
linear model.

\section{Consequences}

Let us now move to the consequences of such an equation of state.  We
shall assume for now on that the macroscopic formalism with the
Lagrangian given by Eq.~(\ref{linear}) is valid to describe a
fermionic carrier cosmic string, provided the string never leaves the
elastic regime. In other words, we shall assume that the string,
whatever its shape, has a curvature radius everywhere much larger than
its thickness and that the Fermi level is below the vacuum mass of the
fermion so that the quantum effects are negligible.

Given a functional relationship between the energy per unit
length $U$ and the tension $T$, one can calculate the
perturbation velocities respectively as\cite{formal}
\begin{equation} c^2_{_T} = {T\over U}\label{ct}
\end{equation}
for the transverse perturbations, and
\begin{equation} c^2_{_L} = -{\hbox{d}T\over \hbox{d}U}\label{cl}
\end{equation}
for the longitudinal ones. In the case at hand~(\ref{UTm2}), this
gives
\begin{equation} c^2_{_L} = 1, \ \ \ \ \ \ \ \ \ \ c^2_{_T} =
{2m^2\over U}-1< 1 \label{clct}
\end{equation}
since $m^2<U<2m^2$ by construction. On a plot $c^2_{_T}$ versus
$c^2_{_L}$, such an equation of state would therefore just be the
line $c^2_{_L}=1$. For cosmological considerations, one may also
consider for instance the back-reaction of the fermions on the
background vortex fields, or even the electromagnetic back
reaction for charged carriers. This means in practice, if one
suppose that these will indeed lead to corrections which in
principle could not be properly placed on such a diagram, that
the corrected equation of state would be a curve somewhere near
the $c^2_{_L}=1$ line.

The situation is exactly the opposite of what happens for a boson
field~\cite{martinpeter} for which it had been found that the
equation of state in this plot is a curve close to the
$c^2_{_T}=1$ line. One can understand this result as a kind of
duality between fermion and boson condensates, the corresponding
equations of state being roughly symmetrical with respect to the
line $c^2_{_L}=c^2_{_T}$. This adds further insight on the fact
that a purely 2-dimensional description is not valid before the
full field theory has been solved. It may be conjectured at this
point that a string carrying a current generated by both fermions
and bosons with an underlying supersymmetric model~\cite{susy}
could produce an equation of state exactly lying on the line
$c^2_{_L}=c^2_{_T}$, i.e. the so-called fixed determinant model
(arising also from a Kaluza-Klein projection~\cite{KK} or as a
smoothed average description of the large scale behavior of a
simple Nambu-Goto model over the small scale
wiggles~\cite{wiggles,integrability}) for which $UT=m^4$. The
advantage of this model, if the conjecture turned out to be a
reasonable approximation of a more realistic equation of state,
lies in its complete integrability~\cite{integrability} in the
case of a flat background. Such a feature might be useful in
network simulations.

The last point that needs be mentioned here concerns vorton
stability. It was shown under rather general conditions that
circular loops reaching an equilibrium state thanks to a current
may suffer from classical instabilities, the fate of which
presumably leading to quantum effects~\cite{mp2}, provided the
equation of state is in the region above the $c^2_{_L}=c^2_{_T}$
line~\cite{vortstab}. Inclusion of the electromagnetic
corrections has also been achieved, showing that these can reduce
the number of vortons that can form during the loops
evolution~\cite{gpb}, but that once they are formed, they are,
classically, more stable~\cite{vortelec}. In our case, if the
corrections do not change drastically the form of the equation of
state, the vortons would exist comfortably below the critical
line. Therefore, we expect them to be much more stable with
respect to classical perturbations. In fact, it is very hard to
imagine anything, except quantum background
interaction~\cite{davis}, that could destabilize a vorton whose
dynamics stems from the Lagrangian~(\ref{linear}).

\section{Conclusion}

Fermionic zero modes trapped in cosmic strings are shown to
follow a vanishing temperature Fermi-Dirac distribution. This is
so because the chirality of the zero modes involved are such that
the only possible interaction of the fermions is through gauge
boson radiation, leading to an effective loss of energy (on
average). As a result, as strings are formed and fermions get
condensed along them in the form of zero modes, populating
arbitrary high energy levels, they have the possibility to decay
radiatively until they reach a zero temperature distribution.
Then, as all other interaction terms are identically vanishing,
they remain in this state which thus happens to be stable.

Assuming therefore such a vanishing temperature fermionic
current-carrying cosmic string, it turns out that the equation of
state relating the energy per unit length $U$ and the tension $T$
is of the self-dual~\cite{carpet1} fixed trace kind, namely
$U+T=2m^2$, with $m$ the characteristic string-forming Higgs
mass. Although fermionic carriers imply the need of more than one
state parameter, this implies that the simplest linear
Lagrangian~(\ref{linear}) provides a good approximation for a
classical description of such a vortex. This could in fact have been
anticipated as this is the only available equation of state that does
not involve any new dimensionfull constant.

Vortons formed with such currents are completely stable, at least
at the classical level (see however Ref.~\cite{davis} for quantum
excitations). Assuming backreaction and electromagnetic
corrections to be small, one finds that the vorton excess
problem~\cite{vortons} is therefore seriously enhanced for
fermionic current-carrier cosmic strings.

\section*{Acknowledgments}

It is a pleasure to thank Brandon Carter, Xavier Martin and Mairi
Sakellariadou for various enlightening discussions.

\end{document}